\documentclass[a4,usenatbib,fleqn]{mn2e}
\usepackage{mathptmx}
\usepackage{amsmath,amssymb,amsbsy}
\usepackage[pdftex]{graphicx}
\usepackage[backref=none]{hyperref}
\usepackage[usenames,dvipsnames]{color}
\usepackage{bm}
\usepackage{txfonts}
\DeclareSymbolFont{cmletters}{OML}{cmm}{m}{it}
\DeclareMathSymbol{v}{\mathalpha}{cmletters}{"76}

\def\rads{\hbox{\rm\hskip.35em rad s}$^{-1}$}

\def\cms{\hbox{\rm\hskip.35em  cm s}$^{-1}$}

\definecolor{MyDarkBlue}{rgb}{0,0.1,0.7}
\hypersetup{pdfborder={0 0 0},colorlinks,breaklinks=true,
  urlcolor={blue},citecolor={MyDarkBlue},linkcolor={magenta}}

\newcommand{\abs}[1]{{\left|#1\right|}}

\newcommand{\apj}{ApJ}
\newcommand{\apjl}{ApJ}

\newcommand{\mnras}{MNRAS}
\newcommand{\nat}{Nature}

\newcommand{\aap}{A{\&}A}
\newcommand{\araa}{ARA{\&}A}

\newcommand{\apss}{Ap{\&}SS}

\newcommand{\prc}{Phys. Rev. C}
\newcommand{\prl}{Phys. Rev. Lett.}

\title[Largest Crab glitch and creep model]{The Largest Crab Glitch and the Vortex Creep Model}

\author[G\"{u}gercino\u{g}lu \& Alpar]{Erbil G\"{u}gercino\u{g}lu$^{1}$\thanks{Contact e-mail: \href{mailto:egugercinoglu@sabanciuniv.edu}{egugercinoglu@sabanciuniv.edu}} and M. Ali Alpar$^{1}$\thanks{Contact e-mail: \href{mailto:alpar@sabanciuniv.edu}{alpar@sabanciuniv.edu}}
\\
$^{1}$Faculty of Engineering and Natural Sciences, Sabanc{\i} University, Orhanl{\i}, Tuzla, 34956 Istanbul, Turkey}

\date{Accepted XXX. Received YYY; in original form ZZZ}

\pubyear{2019}

\begin{document}
\label{firstpage}
\pagerange{\pageref{firstpage}--\pageref{lastpage}}
\maketitle

\begin{abstract}
The Crab pulsar displayed its largest glitch on 2017 November. An extended initial spin-up phase of this largest glitch was resolved, for the first time with high cadence of observations both in radio and X-rays on a time-scale of 2 days. A combination of crustquake and vortex unpinning models is invoked to account for the extended spin-up, magnitude and post-glitch relaxation characteristics of this glitch. We evaluate the extended spin-up followed by the familiar spin-down as due to the creep response to the initial induced inward motion of some vortex lines pinned to broken crustal plates moving inward towards the rotation axis, together with the common and familiar post-glitch creep response to the sudden outward motion of vortices unpinned at the glitch.  Our analysis confirms that the number of unpinned vortices participating in glitches are similar in all Crab glitches, and within an order of magnitude in all glitches from all pulsars. This typical number of unpinned vortices is related to the broken plate size in quakes as triggers for vortex unpinning avalanches. The physical determinant of this universal broken plate size is in turn the critical strain angle in the neutron star crust. Occurrence of this largest Crab glitch after a relatively long inactive period is consistent with accumulation of the pinned vorticity to be tapped.
\end{abstract}





\section{Introduction} \label{sec:intro}

In many pulsars, the regular slow-down of the rotation rate is occasionally interrupted by abrupt spin-up, $\Delta\Omega >0$, a glitch, which is followed by a recovery towards the original state on time-scales from days to months (for the most recent reviews see \citet{espinoza11,yu13,fuentes17}). The fractional glitch magnitudes are in the range $\Delta\Omega /\Omega \sim 10^{-10} - 10^{-5}$. The recovery time-scales following glitches provide strong evidence for the existence of a superfluid component inside the neutron star, since a star consisting of normal matter would relax much faster \citep{baym69}. Glitches are usually accompanied by a jump in the spin-down rate, $\Delta\dot\Omega <0$. The glitch associated $\Delta\Omega$ in the rotation rate does not fully relax back and in many glitches $\Delta\dot\Omega$ is also observed to recover only partially before the next glitch arrives. Observations of glitches and slow post-glitch recoveries can be used to shed light on the various aspects of the physical mechanisms occurring inside neutron stars \citep{haskell15,erbil17b}. 

Superfluids carry angular momentum by sustaining quantized vortex lines. In some parts of the neutron star these vortices can get pinned to local inhomogeneities. Pinning constrains the spin-down of the superfluid. Pulsar glitches are attributed to transfer of angular momentum from interior superfluid components to the crust of a neutron star when metastably pinned vortices are released suddenly \citep{anderson75}. In between glitches superfluid spins down by the average flow of vortices thermally activated against the pinning energy barriers. When a glitch occurs this process, called vortex creep, is affected by the sudden change in rotation rates of the crust and superfluid components. The observed post-glitch relaxation reflects the vortex creep response to glitch induced changes \citep{alpar84}. Another model for glitches is the crustquake model \citep{baym71}. According to this model, the solid crust tends to follow the evolving  equilibrium shape of the star, as it spins down, by suffering successive breaks. These discrete attempts to achieve the equilibrium shape will make the star less oblate, by decreasing the moment of inertia  the rotation rate increases suddenly. Crustquakes alone cannot account for frequent large glitches with $\Delta\Omega /\Omega \sim 10^{-6}$ like those observed in the Vela pulsar. They however can act as triggers for vortex unpinning avalanches \citep{akbal18}. For smaller glitches the rate of glitching may require the joint scenario of crustquake triggered unpinning events \citep{alpar96}.

The Crab pulsar is a young, very fast spinning pulsar and has the largest spin-down rate among all known neutron stars. It is monitored regularly by a dedicated observing program at Jodrell Bank so that its glitches are caught with the occurrence time of the glitch being determined to an accuracy within hours \citep{espinoza14,lyne15}. The Crab pulsar has experienced 30 glitches since its first recorded event in 1969 \footnote{For a continually updated database see Jodrell Bank glitch catalogue at the URL: www.jb.man.ac.uk/pulsar/glitches/gTable.html}. These are typically small glitches with $\Delta\Omega/\Omega\approx10^{-9}$ while occasional large glitches with $\Delta\Omega/\Omega\lesssim 10^{-7}$ are also observed. The minimum resolved glitch magnitude $\Delta\Omega/\Omega\cong 1.7 \times 10^{-9}$ \citep{espinoza14} has placed interesting constraints on the crustal physics \citep{akbal18}. The glitches in the spin-down rate of the Crab pulsar are of the order of $\Delta\dot{\Omega}/\dot{\Omega}\sim 10^{-4}- 10^{-3}$. Part of the glitch associated increment in the spin-down rate, dubbed the `persistent shift' does not relax back by the time of the next glitch \citep{demianski83,wong01,wang12,lyne15}.    

The Crab pulsar shows highly irregular glitch activity \citep{wong01,wang12,lyne15}. Its glitches are not evenly distributed in time although most of them are events of similar size. Crab glitch size distribution obeys a power law while glitch intervals are exponentially distributed \citep{carlin19}. Crab glitched 14 times in a period of 9 years between 1995 and 2004, while since 2011 it entered a period of low activity until its largest glitch on 2017 November \citep{lyne15,shaw18a}. The long term average interval between Crab pulsar's glitches is about 2 years.   

The magnitude $\Delta\Omega/\Omega\sim10^{-9}$ of a typical Crab glitch allows crust breaking at a few year intervals as an energetically viable model in the sense that the expected energy dissipation in crustquakes will not exceed observational bounds. Are the time intervals between glitches compatible with the models? In the vortex creep model the inter-glitch time is given by \citep{alpar84}
\begin{equation}
t_0=\frac{\delta\Omega_{\rm s}}{\left|\dot\Omega\right|},
\end{equation}
where $\delta\Omega_{\rm s}$ is the change in superfluid angular velocity due to vortex discharge. In the crustquake model the glitch repetition time-scale is given by \citep{baym71}
\begin{equation}
t_{\rm q}=\frac{2A(A+B)}{IB\Omega\left|\dot\Omega\right|}\left(\frac{\Delta\Omega}{\Omega}\right),
\end{equation}
where $I$ is moment of inertia of the neutron star, $A$ and $B$ are structural constants related to gravitational binding energy and lattice elasticity, respectively \citep{baym71}. For a typical value 
of $\Delta\Omega/\Omega\approx 10^{-9}$ then  $t_0<1$ yr and $t_{\rm q}\gtrsim 10$ yr are obtained. This shows that neither superfluid vortex unpinning nor crustquake models by themselves can explain Crab glitch intervals, and suggests an interplay between the two processes \citep{alpar96}.
 
In this paper we consider the 2017 November glitch, which exhibited a resolved glitch rise extending over 2 days. This is the largest Crab pulsar glitch so far, larger than the typical Crab glitches by more than an order of magnitude in both the rotation rate and the spin-down rate. \citet{shaw18a} observed this glitch with $\Delta\Omega/\Omega=0.516\times10^{-6}$ and $\Delta\dot\Omega/\dot\Omega=6.969\times10^{-3}$ at MJD 58064 (2017 November 8). \citet{krishnakumar17} observed this glitch with the Ooty Radio Telescope (ORT) around frequency band  326.5 MHz and reported a pulse morphology change concurrent to the glitch. A similar analysis of Jodrell Bank data in two other bands did not confirm such changes, suggesting a process depending on the height in the pulsar magnetosphere \citep{shaw18a}. \citet{zhang18} and Ge et al. (private communication) observed the glitch in X-rays with the Chinese X-ray telescopes X-Ray Pulsar Navigation-I (XPNAV-1) and Insight-HXMT, respectively. The gradual spin-up, extending over about 2 days, was resolved in both radio and X-ray timing with high cadence observations (every few hours). The uncertainty in the actual time of Crab pulsar glitches, in the sense of the time between the last pre-glitch and the first post-glitch fits to the pulsar timing parameters, is of the order of a few hours \citep{espinoza14}. The tightest limit on the rise time of a glitch from any pulsar was observed in one glitch of the Vela pulsar \citep{dodson02} as a step change not resolved below 40 seconds. The 2017 November glitch is the best resolved gradual glitch rise observed to date for the Crab pulsar. The 1989 and 1996 glitches had also shown extended spin-ups with time-scales 0.8 and 0.5 days, respectively \citep{lyne92,wong01} but the glitch rise was sparsely sampled, with only one timing observation during the rise of either glitch. A small glitch with $\Delta\Omega/\Omega= (4.08\pm0.22)\times 10^{-9}$ and $\Delta\dot\Omega/\dot\Omega= (0.46\pm0.11)\times 10^{-3}$ occurred 176 days after the largest glitch of 2017 November \citep{shaw18b}. This small glitch did not affect the recovery after the largest glitch (Mingyu Ge 2018, private communication). 

The 2 day time-scale of the extended spin-up is difficult to be explained either as the time-scale for crust breaking and plate motion or as the time for suddenly unpinned vortices to move radially outward through the superfluid and thereby transfer angular momentum to the crust, which takes less than a minute \citep{graber18}. Explanations for delayed spin-up have been proposed in terms of secular vortex movement as a result of excess heating due to a quake in a hot crust \citep{greenstein79,link96} or a change in the mutual friction strength in a strongly pinned crustal superfluid region as the unpinned vortex front propagates \citep{haskell18}. In the vortex creep model initial extended spin-up of the observed crust will arise due to the response of the vortex creep process to a component of the glitch involving vortex motion inward, towards the rotation axis. Such inward vortex motion can be induced on vortices pinned to a crustal plate which breaks and moves inward to remove crustal stresses. The same event can also lead to unpinning of many more vortices which move rapidly outward, leading to the creep response signatures commonly observed in post-glitch relaxation. A model involving induced inward vortex motion as well as outward motion of unpinned vortices was proposed earlier for the `peculiar' glitch of PSR J1119$-$6127 \citep{weltevrede11,antonopoulou15}. To explain that event \citet{akbal15} assumed that the glitch was triggered by the breaking of a crustal plate. To relieve the stress the broken plate moves towards the rotation axis. Some vortex lines pinned to the broken plate are carried inward, towards the rotation axis. This causes a response of the vortex creep process to contribute an extended post-glitch spin-up of the crust. At the same time, a large number of vortex lines get unpinned in the event and move rapidly outward, in a very short time-scale, yielding the glitch increase in the rotation rate, as usual in the unpinning model for glitches. The vortex creep response to the sudden motion of unpinned vortices causes the subsequent increase in spin-down rate of the observed crust which relaxes on longer time-scales. 

We apply a similar model to analyse the largest Crab glitch. The glitch starts with initial crust breaking and motion of the broken plate, carrying vortices pinned to it inward, while vortices unpinned at the glitch move away from the rotation axis. The creep responses to inward and outward glitch associated vortex motion  contribute to the initial 2 day spin-up and subsequent increased spin-down rate and its relaxation. 
In \S\ref{sec:theory} we briefly review the vortex
creep model. In \S\ref{sec:fit} we apply our extended model to the largest Crab glitch data. We discuss our results in \S\ref{sec:dandc}.
\section{Outline of Theory} \label{sec:theory}

\subsection{The Two Component Model} \label{subsec:twocomp}
A superfluid system achieves rotation by forming quantized vortex lines. It can spin-down by sustaining a flow of these vortex lines in the radially outward direction (away from the rotation axis):
\begin{equation}
\dot\Omega_{\rm s} = -\frac{2\Omega_{\rm s}}{r}v_{\rm r}(\omega).
\label{sfluidsd}
\end{equation}
The radial velocity $v_{\rm r}$ of the vortices depends on the lag $\omega \equiv \Omega_{\rm s} - \Omega_{\rm c}$ between the rotation rate $\Omega_{\rm s}$ of the superfluid and the rotation 
rate $\Omega_{\rm c}$ of  the normal matter crust of the neutron star. The normal matter throughout the neutron star, like electrons and the nuclei forming the crustal solid are rigidly coupled to and co-rotating with the observed outer crust's rotation rate $\Omega_{\rm c}$. The radial vortex flow is set up as the vortices, whose azimuthal motion is with the superfluid at $\Omega_{\rm s}$, interact with ambient normal matter which co-rotates with the crust at $\Omega_{\rm c}$. Hence, $v_{\rm r}$ depends on the lag $\omega$. The functional dependence of $v_{\rm r}$ on $\omega$ depends on the nature of the physical interaction between the vortices and ambient normal matter. If the drag forces between vortices and electrons or phonons are linear in the relative velocity, $v_{\rm r}$ is also linear in $\omega$. For vortex creep processes to be described below, $v_{\rm r}$ is a highly non-linear function of $\omega$. 

Rotational evolution of pulsars is determined by the external torque, which is essentially the dipole radiation torque for an isolated neutron star 
\begin{equation}
I_{\rm c}\dot\Omega_{\rm c} + I_{\rm s}\dot\Omega_{\rm s} = N_{\rm ext},
\label{crusteom}
\end{equation}
where $I_{\rm c}$ and $I_{\rm s}$ are the crustal and superfluid moments of inertia, respectively. Post-glitch relaxation data from pulsars indicate that only a small fraction of the neutron star superfluid is involved in glitches, $I_{\rm s} \lesssim 10^{-2}I_{\rm c}$. This implies that the superfluid core of the neutron star, which contains most of the moment of inertia of the star, is already included in $I_{\rm c}$,  and the superfluid components with smaller $I_{\rm s}$ inferred from the observations that responsible for glitches are all associated with the crustal superfluid. The tight coupling of the core superfluid is explained by electron scattering off the spontaneous magnetization of vortices in the core due to  proton super-currents dragging around them \citep{ALS84}.

Equations (\ref{sfluidsd}) and (\ref{crusteom}) define the two-component model for neutron star dynamics. 
This dynamical system has a steady state solution in which $\dot\Omega_{\rm s} = \dot\Omega_{\rm c} = N_{\rm ext}/I \equiv \dot\Omega_{\rm \infty}$ where $I = I_{\rm c}+I_{\rm s}$ is total moment of inertia. In steady state the lag $\omega_{\rm \infty}$ is constant and can be obtained from Eq.(\ref{sfluidsd}) by setting $\dot\Omega_{\rm s} = \dot\Omega_{\rm \infty}$.
When the force on the vortex lines and therefore $v_{\rm r}(\omega)$ are linear in the lag $\omega$, 
Eq.(\ref{sfluidsd}) assumes the linear form $\dot\Omega_{\rm s} = -\frac{\omega}{\tau}$ where $\tau$ is the relaxation time corresponding to the specified physical interaction between vortex lines and normal matter. The linear response to glitch induced changes is simple exponential relaxation towards steady state \citep{baym69}. The steady state lag is given by $\omega_{\rm \infty} = \abs{\dot\Omega_{\rm \infty}}\tau$.

\subsection{Overview of the Vortex Creep Model} \label{subsec:creep}

In the neutron star crust vortex lines are in a very inhomogeneous medium where it is energetically favourable for vortices to pin to the crystal lattice nuclei with pinning energy gain $E_{\rm p}$. Vortex lines have probabilities proportional to Boltzmann factors to jump between adjacent pinning sites. At any given time and location where some fraction of vortex lines are pinned the superfluid rotation rate will be higher than that of the normal matter which is spinning down with the outer crust under the (external) pulsar torque. The rotational lag $\omega \equiv \Omega_{\rm s} - \Omega_{\rm c}$ introduces a bias on the activation energies for vortices such that vortex motion in the radially outward direction from the rotation axis is energetically favourable. The result will be an average `vortex creep' rate in the radially outward direction which depends exponentially on the lag and allows the superfluid to spin down \citep{alpar84}. Thus, at finite temperature a pinned superfluid is not absolutely pinned in the sense of all vortices remaining pinned all the time. Individual vortices alternatively pin or unpin at different times as they partake in the dynamical process of creep. On this background of creep across the pinning energy landscape, there will be locations where the pinning energies are particularly high and vortices do not take part in creep \citep{cheng88}. Vortices in these pinning traps will be unpinned in catastrophic events when the lag reaches a critical value $\omega_{\rm cr}$. The angular momentum transfer in the pinned superfluid is like charge current in an electric circuit, with vortices as the carriers. Pinning traps are the capacitors, whose discharges constitute the glitches. The continuously ongoing creep is analogous to electric current flow through resistive elements. There is an interplay between glitches and vortex creep. Glitches as sudden vortex discharges reduce the superfluid rotation rate $\Omega_{\rm s}$ and transfer angular momentum to the crust, increasing its rotation rate $\Omega_{\rm c}$. This offsets the lag $\omega$ and can affect creep very strongly, even temporarily stop it, because of the exponential dependence of the creep rate on $\omega$. The effect is observed as a step change in the observed spin-down rate $\dot\Omega_{\rm c}$ which then evolves back towards the pre-glitch situation as the creep process recovers. 

The equation of motion for a vortex line is the Magnus equation, which relates the force $f_{\rm p}$ per unit length of vortex line to the relative velocity of the vortex line with respect to the ambient superfluid. A pinned vortex line moves together with the normal matter, i.e. crust lattice and nuclei to which it is pinned. The critical (maximum) angular velocity lag $\omega_{\rm cr}$ that can be sustained by the maximum available pinning force is
\begin{equation}
f_{\rm p}\; = \frac{E_{\rm p}}{b \; \xi}\;= \; \rho \; \kappa\; R \;\omega_{\rm cr},
\end{equation}
where $R$ is the distance from the rotation axis, $\rho$ is the superfluid density and $\kappa$ is the quantum of vorticity. Here $b$ denotes the spacing between successive pinning sites along the vortex line which can be of the order of the lattice constant or likely longer (much longer) depending on the strength of pinning, and $\xi$ is the radius of the vortex line core, which is of order or larger than the nuclear radius. The vortex line will get unpinned in situations where and when the
lag $\omega$ reaches the critical value $\omega_{\rm cr}$. 

In the vortex creep model, radial vortex velocity is \citep{alpar84}
\begin{equation}
v_{\rm r}=v_{0}\exp \left(-\frac{E_{\rm p}}{kT} \right) \sinh \left( \frac{\omega}{\varpi} \right),
\end{equation}
where $v_{0}\approx 10^{7}$\cms is a microscopic vortex velocity for free (not pinned) vortex  line segments between pinning episodes \citep{alpar84,erbil16} and $\varpi=(kT/E_{\rm p})\omega_{\rm cr}$. 

Depending on the argument of the $\sinh$ function vortex creep has linear or non-linear regimes \citep{alpar89}. In the linear regime vortex creep contributes an exponentially relaxing component to the post-glitch response \citep{alpar89}
\begin{align}
\Delta\dot{\Omega} (t)=-\frac{I_{\rm l}}{I}\frac{\delta\omega}{\tau_{\rm l}}{\rm e}^{-t/\tau_{\rm l}}.
\label{lcreep}
\end{align}
Here $I_{\rm l}$ is the moment of inertia of the linear creep region and its relaxation time is
\begin{equation}
\label{taulin}
\tau_{\rm l} \equiv \frac{kT }{E_{\rm p}} \frac{R \omega_{\rm cr}}{4 \Omega_{\rm s} v_{0}} \exp \left( \frac{E_{\rm p}}{kT} \right),
\end{equation}
where $R$, the distance of the vortex lines from the rotational axis is roughly equal to the neutron star radius. 
The steady state lag for linear creep $\omega_{\rm \infty} = \abs{\dot\Omega_{\rm \infty}}\tau_{\rm l}$ is much less than $\omega_{\rm cr}$. 
Within linear creep regions typically no glitch associated vortex motion takes place, so the glitch induced change in the lag is $\delta\omega = \Delta\Omega$. The linear creep relaxation time $\tau_{\rm l}$ has a very strong dependence on the local pinning energy and internal temperature and can explain relaxation time-scales from several days to a few months. 

In the non-linear creep regime, $\omega_{\infty} \cong \omega_{\rm cr}$ and its contribution with moment of inertia $I_{\rm A}$ to the post-glitch response is \citep{alpar84}
\begin{equation}
\Delta\dot\Omega (t)=-\frac{I_{\rm A}}{I}\abs{\dot\Omega_{\rm \infty}}\left[1-\frac{1}{1+({\rm e}^{t_{0}/\tau_{\rm nl}}-1){\rm e}^{-t/\tau_{\rm nl}}}\right],
\label{ncreep}
\end{equation}
with a non-linear creep relaxation time
\begin{equation}
\tau_{\rm nl}\equiv \frac{kT}{E_{\rm p}}\frac{\omega_{\rm cr}}{\abs{\dot\Omega_{\rm \infty}}}.
\label{taun}
\end{equation}
Creep restarts after a waiting time $t_{0}=\delta\omega/\abs{\dot\Omega_{\rm \infty}}\simeq\delta\Omega_{\rm s}/\abs{\dot\Omega_{\rm \infty}}$. A combined (``stacked") response of many non-linear creep regions yields a power law behaviour, with a constant ``anomalously large" $\ddot{\Omega}$ commonly observed between glitches \citep{yu13}.

Superfluid neutrons will be ``entrained" by the crustal lattice due to Bragg scattering and superfluid coherence effects, thereby acquiring effective masses larger than the bare neutron mass \citep{chamel12}. This leads to a need to employ more superfluid regions complementing the crustal superfluid to model post-glitch relaxation observations. A likely extra creep region beyond the crustal superfluid is the outer core where the neutron superfluid coexists with type II superconducting protons with toroidally oriented quantized flux tubes providing pinning centres for vortex lines \citep{erbil14, erbil17a}. For the Crab pulsar, with the small size of glitches, including even this largest one, the entrainment effect does not necessitate the involvement of more creep regions (the toroidal flux region) beyond the crustal superfluid. However, the presence of a component of relaxation that can be consistently associated with the post-glitch response of toroidal flux regions in all pulsars suggests that such a component is a part of neutron star structure. The expected relaxation time $\tau_{\rm tor}$ for this component is estimated to be $0.5-2$ days for the Crab pulsar \citep{erbil17a} and accounts for the short relaxation times observed after its glitches in the increased glitch activity period quite well \citep{wong01}. The 2017 November glitch is unusual with the presence of the extended spin-up on a similar $\sim2$ day time-scale. It will not be possible to  delineate the post-glitch response of the toroidal flux region from the extended spin-up in this glitch. We shall not include the toroidal flux region's response in our model for the largest glitch.

Under spin-down torques, the vortex motion is predominantly in the radially outward direction and inward vortex movement has very low probability thermodynamically. Inward vortex motion may become possible when an external agent like crustal plate motion due to a quake is operative. This increases the superfluid rotation rate by some $\delta\Omega'_s$ as the vortices pinned to the broken plate are induced to move towards the rotation axis together with the plate$-$when vortices move inward, superfluid rotates faster. The lag $\omega$ thereby increases from its steady state value and creep will be more
efficient than in steady state, ending up with an enhanced vortex creep current in the
radially outward direction. The effect on the crustal rotation rate and its relaxation is obtained from Eq.(\ref{ncreep}) by changing
$t_0$ with $- t'_0$, where $t'_0 \cong \delta\Omega'_{\rm s}/\abs{\dot\Omega_{\infty}}$ describes the offset in non-linear creep due to inward vortex motion. The post-glitch relaxation of creep in an event which involved induced inward vortex motion throughout a superfluid region with moment of inertia $I_{0}$ is \citep{akbal15}:
\begin{equation}
\Delta\dot\Omega=-\frac{I_{0}}{I}\abs{\dot\Omega_{\infty}}
\left[1-\frac{1}{1+({\rm e}^{-t'_0/\tau'_{nl}}-1){\rm e}^{-t/\tau'_{\rm nl}}}\right].
\label{creepin}
\end{equation}
 
A crustquake triggering a glitch is also likely to unpin some vortices which move radially outward and cause the conventional vortex creep response given by Eq.(\ref{ncreep}). In the rare glitches in which induced inward motion of pinned vortices accompanies the outward motion of unpinned vortices, the post-glitch response will contain both of the components given in Eqs.(\ref{ncreep}) and (\ref{creepin}). The response to inward glitch motion has the signature of a gradual initial spin-up as observed after the Crab pulsar's 2017 November glitch. 

\section{Model Fit to the Post-Glitch Timing Data} \label{sec:fit}

The extended glitch and subsequent post-glitch timing data of the largest Crab glitch were obtained by radio timing observations at Jodrell Bank \citep{shaw18a}, by X-ray timing observations with XPNAV-1 satellite \citep{zhang18} and with Insight-HXMT telescope (Ge et al., private communication). The results of Ge et al. consists of 36 timing solution with error bars of which the first 2 data points are in the extended spin-up phase. \citet{shaw18a} obtained prompt post-glitch observations starting 0.6 days after the glitch. The published data of \citet{shaw18a} and \citet{zhang18} extend to about 100 days after the glitch. Observations of Ge et al. started 1.5 days after the glitch and extend to 265 days after the large glitch (including a small glitch which occurred 176 days after the largest glitch). In order to minimise errors and improve the quality of time of arrival  fits by inclusion of more data points, they utilized data from the Nanshan Radio Telescope, Kunming Radio Telescope and  Fermi-GBM/LAT. Their timing solution is consistent with that of \citet{shaw18a} when the two data sets overlap. 
 
\begin{figure}
\centering
\vspace{0.1cm}
\includegraphics[width=1.0\linewidth]{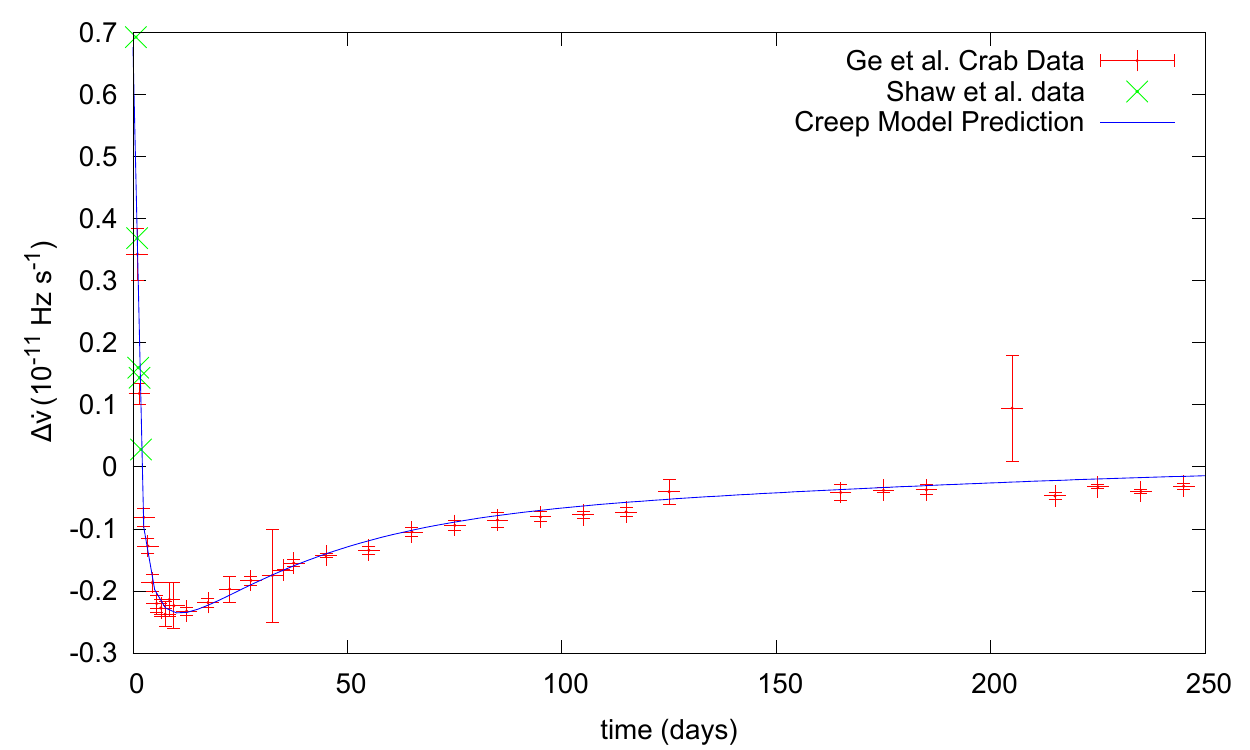}
\caption{Comparison between the Crab pulsar's 2017 November post-glitch data and the creep model (blue). Data with error bars (red) are provided by Ge et al. (private communication). Crosses (green) corresponding to extended spin-up are taken from \citet{shaw18a}. See the online version for the coloured figure.}
\label{datafitcreep}
\end{figure}

High cadence observations of the extended spin-up \citep{shaw18a} and subsequent relaxation (Ge et al. to be submitted) enabled us to analyse the timing behaviour in unprecedented detail. From Eqs.(\ref{lcreep}), (\ref{ncreep}) and (\ref{creepin}) our model is
\begin{align}
\Delta\dot\Omega (t)=
&-\frac{I_{0}}{I}\left|\dot\Omega\right|\left[1-\frac{1}{1+\left(e^{-t'_{0}/\tau'_{\rm nl}}-1\right)e^{-t/\tau'_{\rm nl}}}\right]
\nonumber\\
&-\frac{I_{\rm A}}{I}\left|\dot\Omega\right|\left[1-\frac{1}{1+\left(e^{t_{0}/\tau_{\rm nl}}-1\right)e^{-t/\tau_{\rm nl}}}\right]
-\frac{I_{\rm l}}{I}\frac{\Delta\Omega}{\tau_{\rm l}}e^{-t/\tau_{\rm l}}.
\label{creep}
\end{align}
Comparison between the creep model and the glitch data is displayed in Figure (\ref{datafitcreep}). We plot frequency derivative residuals in terms of  $\Delta\dot\nu\equiv\Delta\dot\Omega/2\pi$ rather than $\Delta\dot \Omega$ in Figure (\ref{datafitcreep}) in order to facilitate comparison with the observational papers. The data point with a large error bar around 205 days, clearly an outlier from the general trend, is presumably due to the artifact of the small glitch which occurred 176 days after the largest glitch. Model parameters are given in Table (\ref{model}). We use the Levenberg$-$Marquardt method in our fit. Errors are included in parenthesis for the entries in the Table (\ref{model}). The three terms in fit function (\ref{creep}) affect different parts of the data on short time-scales and in long term and their contributions can be constrained in this way without contamination among the different exponential relaxation terms. 

\begin{table}
\centering
\caption{Inferred Parameters for model prediction Eq.(\ref{creep}) for data shown in Figure (\ref{datafitcreep}).}
\begin{tabular}{ll}
\hline\hline
Parameter & Value \\
\hline
$\left(\frac{I_{0}}{I}\right) _{-3}$ & 1.50(4) \\
$\left(\frac{I_{A}}{I}\right) _{-3}$ & 2.50(8) \\
$\left(\frac{I_{\rm l}}{I}\right) _{-3}$ & 1.45(5) \\
$\left(\frac{I_{B}}{I}\right) _{-3}$ & 0.6(3) \\
$ t'_{0} (\rm days) $ & 54(4) \\
$ t_{0}(\rm days) $ & 180(10) \\
$ \tau'_{\rm nl} (\rm days) $ & 18(1) \\
$ \tau_{\rm nl} (\rm days) $ & 60(5) \\
$ \tau_{\rm l} (\rm days) $ & 35(3) \\
$ \delta\Omega_{\rm s} (\mbox{\rads})_{-2}$ & 3.6(2) \\
$ \delta\Omega'_{\rm s} (\mbox{\rads})_{-2}$ & 1.08(8) \\
\hline
\label{model}
\end{tabular}
\end{table} 

The first term in Eq.(\ref{creep}) describes the response of vortex creep to the inward vortex motion at the time of the glitch, which {\em increases} the creep rate in that superfluid region above the steady state value. The rate of angular momentum transfer is thereby increased, leading to the extended spin-up of the observed crust. The positive post-glitch frequency deivative residuals associated with this term contribute a quasi-exponential recovery on time-scale $\tau'_{\rm nl}$. The second term in Eq.(\ref{creep}) describes the non-linear creep response from the regions through which the vortices unpinned at the glitch have moved outward rapidly. This type of response is commonly observed after glitches. The sudden motion of the unpinned vortices decreases the lag and thereby effectively stops creep. The effected non-linear creep regions of the superfluid are no longer spinning down. The external torque is now acting on less moment of inertia so that the crust is temporarily spinning down at a slightly higher rate until the lag and the creep rate relax back towards the steady state value. The initial step increase in the magnitude of spin-down rate will recover in a step after a waiting time $t_{0}$ within an interval of order $\tau_{\rm nl}$. The last term describes the response of linear creep regions through which no vortex motion occurred at the time of the glitch. 

\section{Discussion and Conclusions} \label{sec:dandc}

We have analyzed the 2017 November glitch, the largest glitch observed to date in the Crab pulsar, in terms of the vortex creep model. We propose that the glitch is triggered by crust breaking in conjunction with vortex pinning stresses. The broken crustal plates move towards the rotation axis. This is the direction of motion which will relieve the stresses in the solid crust thereby reducing the oblateness of the star to follow the equilibrium shape as the star spins down. The vortices pinned to these plates are induced to move with the plates towards the rotation axis. This leads to extra rapid spin-down of the affected superfluid regions by accelerated vortex creep, and thereby to the extended initial spin-up of the observed pulsar crust immediately following the glitch. 

The crust breaking event also facilitates the unpinning and sudden outward motion of a large number of vortices, larger than the number that remains pinned and moves inward with the broken plate. The sudden outward motion of these unpinned vortices transfers angular momentum to the crust, amplifying the magnitude of the observed glitch in the rotation rate, as usual in the vortex unpinning model for glitches. This results in a total increase in the crust rotation rate, up from the contribution of the triggering crust breaking event. The subsequent components of post-glitch relaxation are interpreted as the non-linear creep response of the crustal superfluid to the radially outward motion of the unpinned vortices, in the usual manner of the creep model for most glitches. In addition to these a third component of post-glitch response, a simple exponential relaxation, is contributed by the linear creep regions of the crustal superfluid in which no glitch associated vortex motion has taken place. The two previous relatively large glitches of the Crab pulsar which also displayed extended spin-ups \citep{lyne92,wong01} probably involved a similar mechanism. Recent reanalysis of the 2004 and 2011 Crab glitches by \citet{ge19} revealed that these large glitches also showed extended spin-ups with timescales 1.7(8) days and 1.3(3) days, respectively. These observations strengthen our idea that large glitches start by quakes which act as triggers for large scale vortex unpinning avalanches. Computer simulations by \citet{warszawski11} confirmed that for vortex lines to unpin in a catastrophic manner and initiate avalanches these vortices should be very close to the threshold for unpinning. In the nonlinear regime of the vortex creep steady state lag is very close to the critical lag for unpinning (see Eq.(\ref{lagunpin}) below) and crustquakes act as trigger for vortex unpinning avalanches.

In addition to the changes in the spin-down rate, one must also take into account the angular momentum balance at the time of the glitch, which gives 
\begin{equation}
\frac{\Delta\Omega}{\Omega}=\left(\frac{I_{\rm A}}{I}+\frac{I_{\rm B}}{I}\right)\frac{\delta\Omega_{\rm s}}{\Omega_{\rm c}}-\left(\frac{I_{0}}{I}\right)\frac{\delta\Omega'_{\rm s}}{\Omega_{\rm c}}-\frac{\Delta I}{I},
\end{equation}
where $I_{\rm B}$ is moment of inertia of vortex free (capacitor) regions which do not sustain creep and therefore do not contribute to the spin-down rate but do contribute to angular momentum transfer as unpinned vortices move through. The reduction in moment of inertia due to crust breaking gives a negligible contribution $\abs{\Delta I/I} \sim 10^{-9}$. The inferred model parameters are listed in Table (\ref{model}).
The moments of inertia $I_{\rm A}$ and $I_{\rm B}$ are associated with non-linear creep regions and vortex free regions, respectively. $I_{0}$ is the moment of inertia associated with non-linear creep response to the induced inward motion of vortices pinned to the broken plate. While $I_{\rm l}$ is the moment of inertia of the crustal superfluid region associated with the linear creep. The total moment of inertia fraction of the creep regions in the neutron star crustal superfluid involved in the glitch is $I_{\rm creep}/I=(I_{\rm A}+I_{\rm B}+I_{0}+I_{\rm l})/I=6.07 \times10^{-3}$. Even if the entrainment effect \citep{andersson12,chamel13} is taken into account the inferred $I_{\rm creep}$ participating in the glitch event remains well within the moment of inertia of the crustal superfluid regions. For the Crab pulsar, the observed changes in the spin-down rate and the inferred moments of inertia fractions participating in the glitch (see Table (\ref{model})) are an order of magnitude smaller than those of the Vela. The creep response of the toroidal flux region is therefore not required in order to accommodate the entrainment effect. Nevertheless if present in Vela and other pulsars, the toroidal region's contribution should also be present in the Crab pulsar. Indeed we find that the $\tau_{\rm tor} \lesssim 2$ days value evaluated for Crab pulsar parameters and moment of inertia fraction for the toroidal region can be identified with one exponentially relaxing component, with $\tau \cong 2$ days in the Crab pulsar postglitch timing.

The inward transported vortices due to crustal plate motion in a quake promptly migrate outwards in the accelerated creep response. These vortex lines are responsible for the extended spin-up process. The resulting initial extended spin up contribution to the glitch magnitude is
\begin{equation}
\Delta\nu_{0}\simeq\frac{I_{0}}{2I}|\dot\nu|t'_{0}\sim 1.29\times10^{-6} \mbox{Hz},
\end{equation}
according to the model fit. This is consistent with the observed value $\Delta\nu_{0}=1.07\times10^{-6} \mbox{Hz}$ \citep{shaw18a}.

The broken plate size $D$ in a crustquake can be determined from simple geometrical arguments \citep{akbal15}:
\begin{align}
I_{0}/I & \cong \frac{4 \pi \rho_{s}  R^{4} D \sin{\alpha} \cos^{2}{\alpha}  }{(2/5)M R^{2}}
\simeq (15/2) \sin{\alpha} \cos^{2}{\alpha}~(D/R) \nonumber\\ &\sim(0.8-2.7)\frac{D}{R},
\end{align}
assuming a uniform density neutron star, and adopting the range of inclination angle $\alpha
\cong 45^{\circ}-70^{\circ}$ indicated by radio polarization studies \citep{watters09, du12}.
Using the value of $I_0/I $ obtained from our model fits (from Table (\ref{model})) one obtains $D\;=(5.7-18.1)R_{6}$ m where $R_{6}$ denotes neutron star radius in units of $10^{6}$ cm. The number of vortices pinned to one broken plate and induced to move inwards with it is 
\begin{equation}
 \delta N_{\rm plate} \sim D^{2} \frac{2\Omega}{\kappa}
\cong (5.97\times10^{10}-6.08 \times 10^{11})R_{6}^{2},
\end{equation}
while the total number of pinned vortices which moved inward with all the broken plates is related to the increase $\delta\Omega'_{\rm s}$ in superfluid rotation rate due to inward vortex motion. Using the value of $\delta\Omega'_{\rm s}$ from our fits (from Table (\ref{model})) we obtain
\begin{equation}
 \delta N_{\rm inward} =2 \pi R^{2}\frac{\delta\Omega'_{\rm s}}{\kappa}\simeq3.39\times10^{13}.
\end{equation}
The number of plates involved in the glitch is obtained as $N_{\rm inward}/N_{\rm plate}\sim10^{3}$, in
agreement with the estimated number of plates in a broken ring, defined as the locus of similar critical stresses, at radius $\sim R$ from the rotation axis. This number of plates is $\sim 2\pi R/D\sim 10^{3}$.
The number of vortices which unpinned and moved outward thereby amplifying the glitch to the observed total magnitude can be obtained likewise from the decrease $\delta\Omega_{\rm s}$ in superfluid rotation rate due to the outward vortex motion:
\begin{equation}
 \delta N_{\rm outward} =2 \pi R^{2}\frac{\delta\Omega_{\rm s}}{\kappa}\simeq1.13\times10^{14}.
\end{equation}
These vortex numbers are of the same order as inferred for other glitches of the Crab \citep{alpar96,akbal18} and Vela pulsars \citep{akbal17} as well as in PSR J1119$-$6127 \citep{akbal15}. Statistical studies show that the associated ratio $\delta\Omega_{\rm s}/\Omega_{\rm s}$ also seems to lie in a common range in all pulsars irrespective of age  \citep{alpar94}. The regularity in the number of vortices involved in glitches of very different magnitude $\Delta\Omega/\Omega$ can be explained by the notion that all these glitches are being triggered by the breaking of crustal plates of similar size, leading to unpinning of similar number of vortices. The similarity in broken plate sizes is in turn due to a fundamental physical property of the neutron star crust, namely the critical strain angle $\theta_{\rm cr} \sim 10^{-1}$ \citep{horowitz09,baiko18}, which sets the broken plate size in terms of crust thickness as $D/\ell_{\rm crust} \sim \theta_{\rm cr}$ . This large value of the critical strain angle reflects the lack of screening by the highly relativistic electrons in the neutron star crust \citep{alpar85}. The glitch magnitude is then proportional to the radial distance through which the unpinned vortices move, which is smaller in young pulsars but can involve the entire crustal superfluid in the Vela and older pulsars. Thus, while the number of unpinned vortices involved in glitches is in a universal range set fundamentally by the critical strain angle, the large difference in glitch sizes reflects the range in the distances travelled (superfluid moments of inertia affected) in the sudden radially outward motion of unpinned vortices.

The post-glitch relaxation of this largest Crab glitch involves responses of various superfluid layers with different time-scales. Our model Eq.(\ref{creep}) involves five time-scales, displayed in Table (\ref{model}). Of these $\tau_{\rm nl}=60$ days and $\tau'_{\rm nl}=18$ days are, respectively, associated with the response of superfluid regions where outward and inward motion of vortices take place during the glitch. The combined  response fits the initial extended spin up phase of $\simeq2$ days. The creep offset time-scales $t_{0}=180$ days and $t'_{0}=54$ days are determined by the number of vortices transported outward and inward in the glitch, respectively. The non-linear creep response to inward vortex motion, the first term in Eq.(\ref{creep}) is a quasi-exponential relaxation. By contrast, the non-linear creep response to outward vortex motion, the second term in Eq.(\ref{creep}) describes a recovery of the initial step in spin-down rate at the waiting time $t_{0}$. As creep returns to steady state on this time-scale it gives a rough estimate for the time of the next glitch \citep{alpar84}. In this connection it is interesting that the subsequent glitch was observed 176 days after the largest glitch \citep{shaw18b}. The third component, describing the response of a linear creep region is in the form of exponential relaxation with $\tau_{\rm l}=35$ days.  

A certain amount of energy is dissipated as glitch associated changes occur in various components of the neutron star while the total angular momentum is conserved. The energy dissipation due to a quake (crust breaking) is given by \citep{baym71,alpar83}
\begin{equation}
\Delta E_{\rm elastic}=\frac{1}{2}I_{\rm c}\Omega_{\rm c}^2\abs{\frac{\Delta I}{I}}.
\end{equation}
With $\Delta I / I \sim 10^{-9}$ due to crust breaking the dissipated elastic energy amounts to $\lesssim 10^{40}$ ergs. The sudden vortex motion in the glitch also dissipates energy. The amount of rotational kinetic energy dissipated is
\begin{equation}
\Delta E_{\rm rot}= (I_{\rm A} \delta \Omega_{\rm s}+I_{0}\delta \Omega'_{\rm s})\omega_{\infty} \cong (I_{\rm A} \delta \Omega_{\rm s}+I_{0}\delta \Omega'_{\rm s})\omega_{\rm cr} \lesssim 10^{40}\:{\rm ergs},
\end{equation}
where we have included inward and outward vortex motion, used $\omega_{\rm cr} \sim 10^{-1}$\rads \citep{erbil16} and taken the total moment of inertia of the neutron star to be $I \sim 10^{45} $ gm cm$^2$. Thus, the total amount of energy dissipation is $\Delta E_{\rm tot}\lesssim 10^{40}$ ergs.

If this energy release is radiated entirely in photons from the neutron star surface (i.e. not allowing for neutrino losses from the interior) on a time-scale $\Delta t $ the corresponding surface black body temperature is
\begin{equation}
T=\left(\frac{\Delta E_{\rm tot}}{4\pi \sigma R^{2}\Delta t }\right)^{1/4} 
\cong 3.5 \times 10^6 \left(\frac{\Delta E_{\rm tot, 40}}{R_6^{2}\Delta t(\rm d) }\right)^{1/4} \mbox{K},
\end{equation}
where $\Delta E_{\rm tot, 40}$ denotes the dissipated energy in units of $10^{40}$ergs, $R_6$ is the neutron star's radius in $10^6$ cm and $\Delta t (\rm d)$ is the duration, in days, for thermal diffusion to heat the surface. The current surface black body temperature upper limit for the Crab pulsar is $T_{\rm s}<2.6\times10^{6}$ K \citep{yakovlev04}. To cause a detectable change in the observed surface temperature\ the dissipated energy in the inner crust should arrive at the surface as a pulse in temperature of duration $ \lesssim 3$ days. This has not been detected in the present glitch or earlier Crab glitches \citep{tang01,lyne15,shaw18a,zhang18}. The thermal diffusion time-scale is indeed estimated to be of the order of a few days or longer \citep{gnedin01}. The energy dissipation in glitches of the magnitude of those observed from the Crab pulsar does not violate observational bounds on surface thermal signals associated with the glitch. 

The glitch activity $A_{\rm g}\leq (I_{\rm s}/I)(\abs{\dot\Omega}/\Omega)$ \citep{link99} or the rate of glitches \citep{alpar94} are expected to be proportional to the spin-down rate. Although the Crab has the highest spin-down rate amongst pulsars, it has an exceptionally low glitch rate \citep{fuentes17}. \citet{alpar96} suggested that the hot crust relieves its stress by plastic flow in the interglitch period so that large and frequent glitches do not occur for this pulsar. 

The maximum expected glitch magnitude for the Crab pulsar after the inactive period may be estimated as follows. Between 2011 and 2017 ($t_{\rm g}=2189$ days) accumulated reservoir superfluid angular velocity is $\delta\Omega_{\rm s}=t_{\rm g}\left|\dot\Omega \right| \cong 0.47$\rads. The fractional moment of inertia of the  decoupled part of the crustal superfluid since the previous glitch is determined from the permanent increase in the spin-down rate residual offset from the previous glitch as $I_{\rm s}/I=\Delta\dot\nu_{\rm p}/\dot\nu=3.32\times10^{-4}$. Then the expected maximum glitch magnitude that can be attained in this period is
\begin{equation}
\frac{\Delta\Omega}{\Omega}=\frac{I_{\rm s}}{I}\frac{\delta\Omega_{\rm s}}{\Omega}\cong 8.2\times10^{-7},
\end{equation}
which agrees with the observed largest Crab glitch $\Delta\Omega/\Omega=5.16\times10^{-7}$. This suggests that in the period of low activity the Crab stored almost all angular momentum stress without relieving a significant part of it. This may be due to cooler crust in the lack of heating sources of quakes and superfluid friction associated with glitches in that period. The difference between the critical lag for vortex unpinning and steady state angular velocity lag is given by
 \citep{alpar89,erbil16}
\begin{equation}
\omega_{\rm cr}-\omega_{\infty}=\omega_{\rm cr}\left(\frac{kT}{E_{\rm p}}\right)\ln\left(\frac{2v_{0}\Omega_{\rm s}}{\left|\dot\Omega \right|R}\right),
\label{lagunpin}
\end{equation}
In a period of high glitch activity higher temperatures can be maintained due to energy dissipation associated with glitches in regions of the crust where the glitches originate. This may lead to an eventual increase in the temperature dependent critical conditions for glitches and to an inactive period without glitches allowing for an eventual large glitch \citep{erbil16}. This agrees with the occurrence of the largest Crab glitch after a relatively long period without glitch activity.

\section*{Acknowledgements}
\addcontentsline{toc}{section}{Acknowledgements}

We thank Dr. Mingyu Ge and Prof. Shuang-Nan Zhang for kindly providing us with unpublished timing solutions, and for useful discussions and correspondence. We acknowledge the referee for helpful comments. This work is supported by the Scientific and Technological Research Council of Turkey
(T\"{U}B\.{I}TAK) under the grant 117F330. M.A.A. is a member of the Science Academy
(Bilim Akademisi), Turkey.





\bsp	
\label{lastpage}
\end{document}